\documentclass[12pt]{article}

\usepackage{scicite}
\usepackage{times}
\usepackage{graphicx}
\usepackage{caption}
\usepackage{multirow}
\usepackage{xcolor}
\usepackage{hyperref}

\title{Modeling Engagement in Long-Term, In-Home Socially Assistive Robot Interventions for Children with Autism Spectrum Disorders} 

\author
{Shomik Jain$^{1\ast}$, Balasubramanian Thiagarajan$^{1}$, Zhonghao Shi$^{1}$, \\Caitlyn Clabaugh$^{1}$, Maja J. Matari\'{c}$^{1\ast}$\\
\\
\normalsize{$^{1}$Interaction Lab, University of Southern California}\\
\normalsize{Los Angeles, CA 90089, USA}\\
\\
\normalsize{$^\ast$To whom correspondence should be addressed; Emails: shomikja@usc.edu, mataric@usc.edu.}\\
}

\date{}

\newcommand{\beginsupplement}{%
        \setcounter{table}{0}
        \renewcommand{\thetable}{S\arabic{table}}%
        \setcounter{figure}{0}
        \renewcommand{\thefigure}{S\arabic{figure}}%
     }

\begin{document} 

\captionsetup[figure]{labelfont={bf},name={Fig.},labelsep=period}
\captionsetup[table]{labelfont={bf},name={Table},labelsep=period}

\baselineskip24pt

\maketitle 

\hspace{3cm}
\hspace{3cm}
\begin{center}
\color{red} \noindent 
This manuscript was published in Science Robotics on February 26, 2020. This version has not undergone final editing. Please refer to the complete version of record at \href{https://robotics.sciencemag.org/content/5/39/eaaz3791}{\textit{https://robotics.sciencemag.org/content/5/39/eaaz3791}}. The manuscript may not be reproduced or used in any manner that does not fall within the fair use provisions of the Copyright Act without the prior, written permission of the American Association for the Advancement of Science (AAAS).
\end{center}

\newpage
\section*{Abstract}
\textbf{Socially assistive robotics (SAR) has great potential to provide accessible, affordable, and personalized therapeutic interventions for children with autism spectrum disorders (ASD). However, human-robot interaction (HRI) methods are still limited in their ability to autonomously recognize and respond to behavioral cues, especially in atypical users and everyday settings. This work applies supervised machine learning algorithms to model user engagement in the context of long-term, in-home SAR interventions for children with ASD. Specifically, we present two types of engagement models for each user: (i) generalized models trained on data from different users; and (ii) individualized models trained on an early subset of the user's data. The models achieved approximately 90\% accuracy (AUROC) for post hoc binary classification of engagement, despite the high variance in data observed across users, sessions, and engagement states. Moreover, temporal patterns in model predictions could be used to reliably initiate re-engagement actions at appropriate times. These results validate the feasibility and challenges of recognition and response to user disengagement in long-term, real-world HRI settings. The contributions of this work also inform the design of engaging and personalized HRI, especially for the ASD community.}

\newpage
\section*{Introduction}

{\it Socially assistive robotics (SAR)} is a promising new subfield of human-robot interaction (HRI), with a focus on developing intelligent robots that provide assistance through social interaction \cite{feil_defining, mataric_sar_handbook}. As overviewed in this journal \cite{mataric_sar_science}, researchers have been exploring SAR as a means of providing accessible, affordable, and personalized interventions to complement human care. However, HRI methods are still limited in their ability to autonomously perceive, interpret, and naturally respond to behavioral cues from atypical users in everyday contexts. This hinders the ability of SAR interventions to be tailored toward the specific needs of each user \cite{rudovic_science, mutlu_behavior}. 

These HRI challenges are not only amplified in the context of SAR for individuals with {\it autism spectrum disorders (ASD)}, but ASD is also the context where SAR is especially promising. ASD is a developmental disability characterized by difficulties in social communication and interaction. About 1 in 160 children worldwide are diagnosed with ASD \cite{who_autism}, with a higher rate of 1 in 59 children in the United States \cite{cdc_autism}. Therapists offer individualized services for helping children with ASD to develop social skills through games or storytelling \cite{ospina_interventions}, but such services are not universally accessible or affordable. To this end, researchers have been actively exploring SAR for children with ASD \cite{scassellati_matric_review}. Several short-term studies have already shown SAR to support learning in ASD users \cite{diehl_review}. Moreover, in this journal, Scassellati et al. \cite{scassellati_science} reported on a long-term, in-home SAR intervention that helped children with ASD to improve social skills such as perspective-taking and joint attention with adults.

\begin{figure}[t!] 
    \centering
    \includegraphics[width=0.6\linewidth]{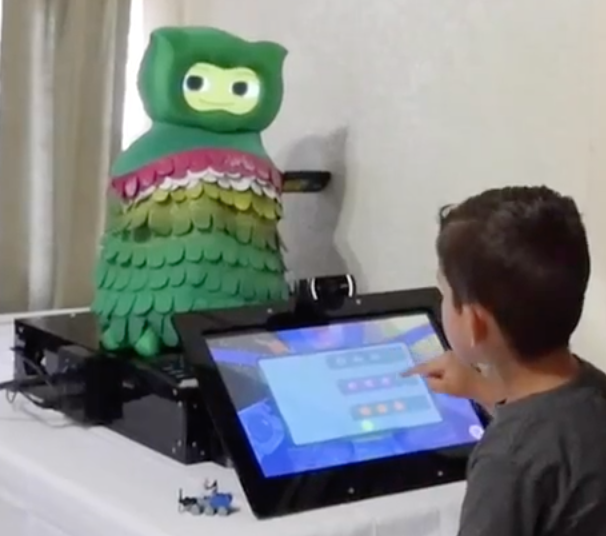}
    \caption{\bf{Long-term, real-world SAR intervention setup.} \normalfont{In this month-long, in-home study, child participants with ASD played math games on a touchscreen tablet, while a socially assistive robot used multimodal data to provide personalized feedback and instruction \cite{scienceNation_video}.}} 
    \label{interventions}
\end{figure}

SAR systems must engage users to be effective. {\it Robot perception of user engagement} is a key HRI capability that makes it possible for the robot to achieve the specified goals of the interaction and, in the SAR context, the intervention  \cite{esteban_rulebased}. Past work has used rule-based methods to approximate the engagement of children with ASD. For instance, Kim et al. \cite{kim_emotion} indirectly assessed engagement by estimating emotional states from audio data. Esteban et al. \cite{esteban_rulebased} gauged engagement by using the frequency of measured variables, such as how many times the child looked at the robot. More recently in this journal, Rudovic et al. \cite{rudovic_science} used supervised machine learning (ML) to model engagement in a single-session laboratory study. The study focused on developing post hoc personalized models using deep neural networks, and achieved an average agreement of 60\% with human annotations of engagement on a continuous scale from -1 to +1.

{\it This article addresses the feasibility and challenges of applying supervised ML methods to model user engagement in long-term, real-world human-robot interactions, with a focus on SAR interventions for children with ASD.} The contributions of this work differ from past work in two key aspects. 

First, the methods and results are based on data from month-long, in-home SAR interventions with 7 child participants with ASD. Whereas single-session and short-term studies of SAR for ASD are numerous \cite{clabaugh_shortterm}, the work by Scassellati et al. \cite{scassellati_science} in this journal is the only other long-term, in-home SAR for ASD study conducted to date. Long-term, in-home studies and data collections are important for many reasons: They more realistically represent real-world learning environments, they provide more opportunities for the user to learn and interact with the robot, and they produce more relevant training datasets \cite{dautenhahn_real_world}. Furthermore, long-term, in-home settings present new modeling challenges, given the significantly larger quantity and variance in user data. 

Second, this work emphasizes engagement models that are practical for online use in human-robot interactions. With a supervised ML approach, models require labeled data for training, which are often expensive or unfeasible to obtain. Previous works report on models trained and tested on randomly sampled subsets of each participant's data \cite{rudovic_science}. However, that approach is impractical for online use if labeled training data for a given user are obtained chronologically after the testing data. In contrast, this work presents, for each user: (i) \textit{generalized models} trained on data from different users; and (ii) \textit{individualized models} trained on an early subset of the user's data. As detailed in Materials and Methods, models were developed for different numbers of training users in generalized models and varying sizes of early subsets in individualized models. An {\it early subset of the data} is defined as the first $X$\% of a user's data sorted chronologically. Furthermore, this work also analyzes the temporal structure of model predictions to examine the possibility of initiating re-engagement actions at appropriate times.  

The presented engagement models are trained on data from month-long, in-home SAR interventions. During interventions, child participants with ASD played space-themed math games on a touchscreen tablet while a Stewart platform robot named Kiwi provided verbal and expressive feedback \cite{short_cordial}. The robot's feedback and instruction challenge levels were personalized to each user's unique learning patterns with reinforcement learning over the month-long intervention. All participants showed improvements in reasoning skills and long-term retention of intervention content \cite{clabaugh_iser, mahajan_frontiers}. Over the month-long data collection, we collected an average of 3 hours of multimodal data across multiple sessions for each participant, including video, audio, and performance on the games. As Figure \ref{interventions} shows, a USB camera mounted on top of the game tablet recorded a front view of the user. Visual and audio features were extracted from the camera data and performance features were derived from the answers to game questions recorded on the tablet. As detailed in Materials and Methods, the open-source data processing tools used to extract these features are appropriate for online use in HRI contexts \cite{baltrusaitis_openface, cao_openpose, boersma_praat}. 

This work frames engagement modeling as a \textit{binary classification problem}, similar to most previous relevant works \cite{rudovic_science}. Participants were annotated as engaged or disengaged in each camera frame using standard definitions of engagement as a combination of behavioral, affective, and cognitive constructs \cite{scassellati_matric_review}. A participant was considered to be engaged when paying full attention to the interaction, immediately responding to the robot's prompts, or seeking further guidance from others in the room. The binary labels simplify the representation of participants' behavior, which may vary in degree of engagement and disengagement \cite{youssef_engagementlevels}. However, temporal patterns in binary labels can provide additional context \cite{clabaugh_iser}; therefore, this work also analyzes the length of time a participant is continuously engaged and disengaged. Trained annotators labeled engagement, and an inter-rater reliability of $k=0.84$ (Fleiss' Kappa) was achieved. The Materials and Methods section provides additional details about the data and the annotation process. 

This article focuses on post hoc models of user engagement based on data from month-long, in-home SAR interventions. The presented approaches are suitable for online perception of engagement, and are intended to inform the design of more engaging and personalized HRI. The contributions of this work especially aim to improve SAR's effectiveness in supporting learning by children with ASD. 

\section*{Results}

This work presents two types of supervised ML models of user engagement in long-term, real-world HRI intended for online implementation: (i) generalized models trained on data from different users; and (ii) individualized models trained on data from early subsets of the users' interventions. On average, these models achieved area under the receiver operating characteristic (AUROC) values of approximately 90\%. {\it AUROC} is a commonly-used ML metric for binary classification problems; specifically, it measures the probability that the models would rank a randomly chosen engaged instance higher than a randomly chosen disengaged instance \cite{bradley_auroc}. In order to evaluate these two approaches, we also implemented models trained on random samples of all user data. Random sampling yielded significantly higher recall for disengagement compared to generalized and individualized models. This is likely because the month-long, in-home setting led to a large variance in both engagement states and recorded data. Variance in data manifested not only across participants but also within each participant, highlighting an important characteristic of real-world HRI in the ASD context. Despite the lower recall and higher variance for disengagement, temporal patterns in model predictions can be used to reliably initiate re-engagement actions at appropriate times.

\paragraph*{Observed User Engagement\newline}

\begin{figure}[t!] 
    \centering
    \includegraphics[width=\linewidth]{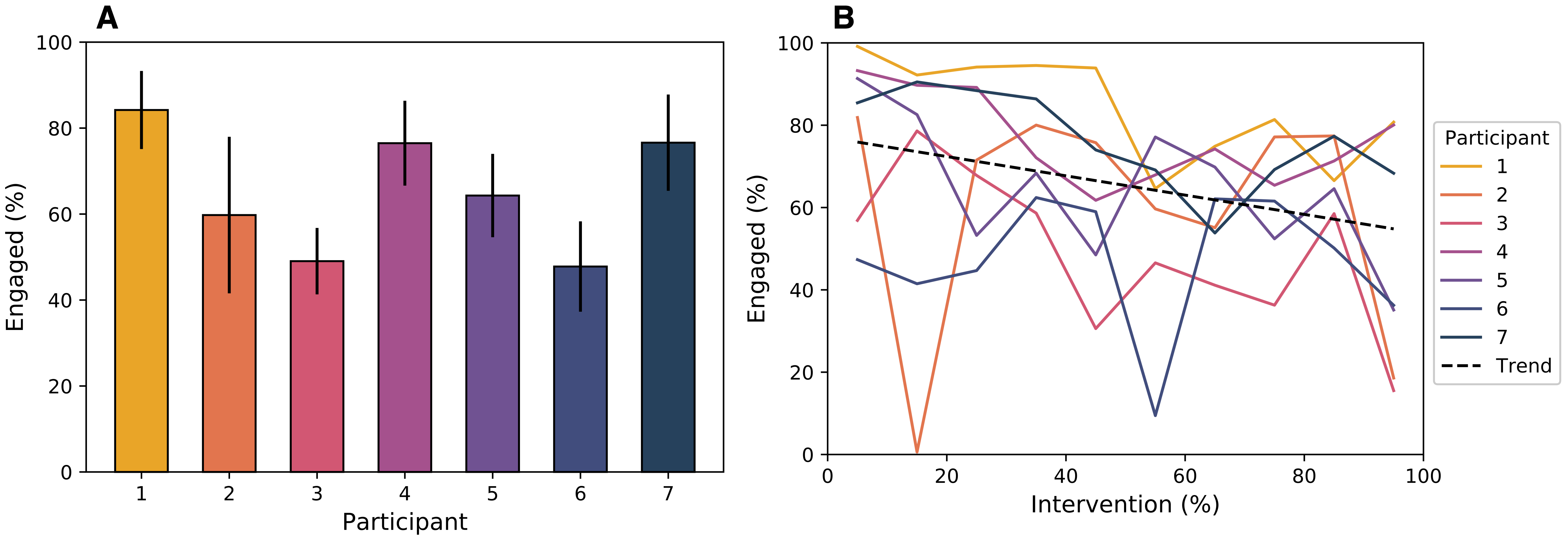}
    \caption{\bf{Engagement by participant.}
    \normalfont{A significant variance in engagement was observed across participants (}\bf{A}\normalfont{) and for each participant (}\bf{B}\normalfont{). A decreasing trend ($p<0.01$) in engagement was also observed over the month-long intervention (}\bf{B}\normalfont{), indicating the need for online engagement recognition and response.}
    } 
    \label{observed_engagement}
\end{figure}

Over the course of the month-long, in-home intervention, participants were engaged an average of 65\% of the time during the child-robot interactions. However, engagement varied considerably across participants and for each participant, as shown in Figure \ref{observed_engagement}. Average engagement for participants ranged from 48\% to 84\%, with a standard deviation of 14\%. Analyzing each participant's engagement chronologically over 10\% increments also showed a standard deviation of 15\%. Moreover, all participants had a significant ($p<0.01$) decrease in engagement over the month-long intervention, as determined by a regression t-test and shown by the plotted trend line. For example, Participant 2 was engaged 82\% of the time in the first 10\% and only 19\% of the time in the last 10\% of the month-long intervention. 

This substantial variance in user engagement over the course of a long-term, real-world study indicates the need for online recognition of and response to disengagement. This study observed higher engagement for all participants shortly after the robot had spoken. Specifically, participants were engaged about 70\% of the time when the robot had spoken in the previous minute, but less than 50\% of the time when the robot had not spoken for over a minute. This validates the use of appropriately-timed robot speech as a tool for eliciting and maintaining user engagement.

\paragraph*{Generalized and Individualized Model Results\newline}

This work presents generalized and individualized models of user engagement using data from long-term, in-home SAR interventions. As detailed in Materials and Methods, generalized models were developed by training on data from a given subset of users and then testing on different users. Individualized models were developed by sorting a user's data chronologically and using an early subset for training and later subset for testing the model. We designed these two approaches to be feasible for online use in HRI; the labeled data required for supervised ML models are practical to obtain in both cases. In order to evaluate these approaches, models trained on random samples of all users' data were also implemented, despite random sampling not being feasible for online use. 

\begin{figure}[t!] 
    \centering
    \includegraphics[width=\linewidth]{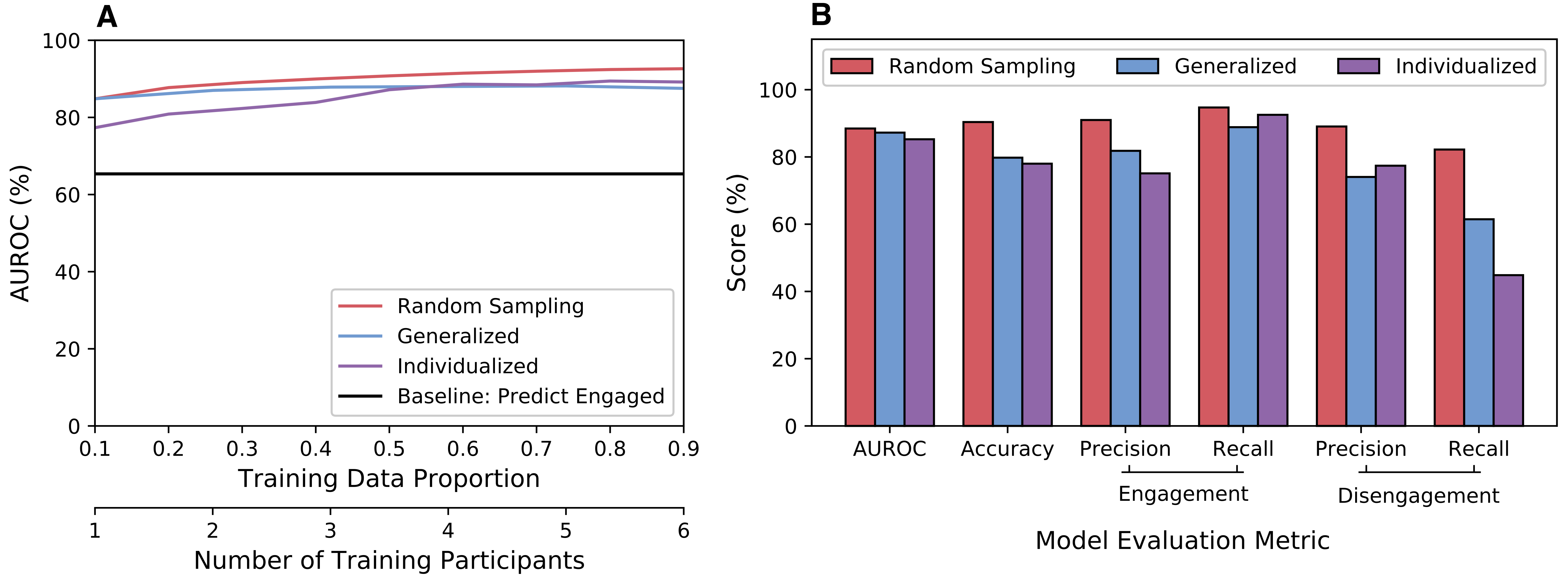}
    \caption{\bf{Model results.}
    \normalfont{Generalized models trained on different users and individualized models trained on early subsets of the intervention achieved comparable AUROC to models trained on random samples of all users' data (}\bf{A}\normalfont{), but had much lower recall for disengagement (}\bf{B}\normalfont{).}} 
    \label{model_results}
\end{figure}

As shown in Figure \ref{model_results}A, generalized and individualized models achieved approximately 90\% AUROC. For generalized models, the number of training users had little effect on AUROC; models trained on six users resulted in only a 3\% improvement in AUROC over models trained on one user. On the other hand, individualized models performed better with additional data; models trained on the first 10\% of a user's data only achieved 77\% AUROC, whereas models trained on the first 50\% of a user's data achieved 87\% AUROC. Overall, both generalized and individualized models achieved comparable AUROC to models trained on random samples, which obtained 90\% AUROC by training on as little as 30\% of data across all users.  

However, generalized and individualized models differ from models trained on random samples when considering other ML evaluation metrics such as precision and recall. For a given class (engagement or disengagement), {\it precision} measures the proportion of predictions of the class that are correct, and {\it recall} measures the proportion of actual instances of the class that are predicted correctly. As Figure \ref{model_results}B shows, there is an especially large difference between models in recall for disengagement. On average, training on random samples resulted in 82\% recall for disengagement, whereas training on different users and early subsets resulted in only 61\% and 45\% recall, respectively. This indicates that generalized and individualized models would produce a high number of false negatives for detecting disengagement if implemented online in HRI. Supplementary Tables \ref{sup_gen_results}, \ref{sup_ind_results}, and \ref{sup_rand_results} contain detailed model results for all approaches and evaluation metrics. 

\paragraph*{Variance in Data Across Users, Sessions, and Engagement States\newline}

\begin{figure}[h!] 
    \centering
    \includegraphics[width=0.95\linewidth]{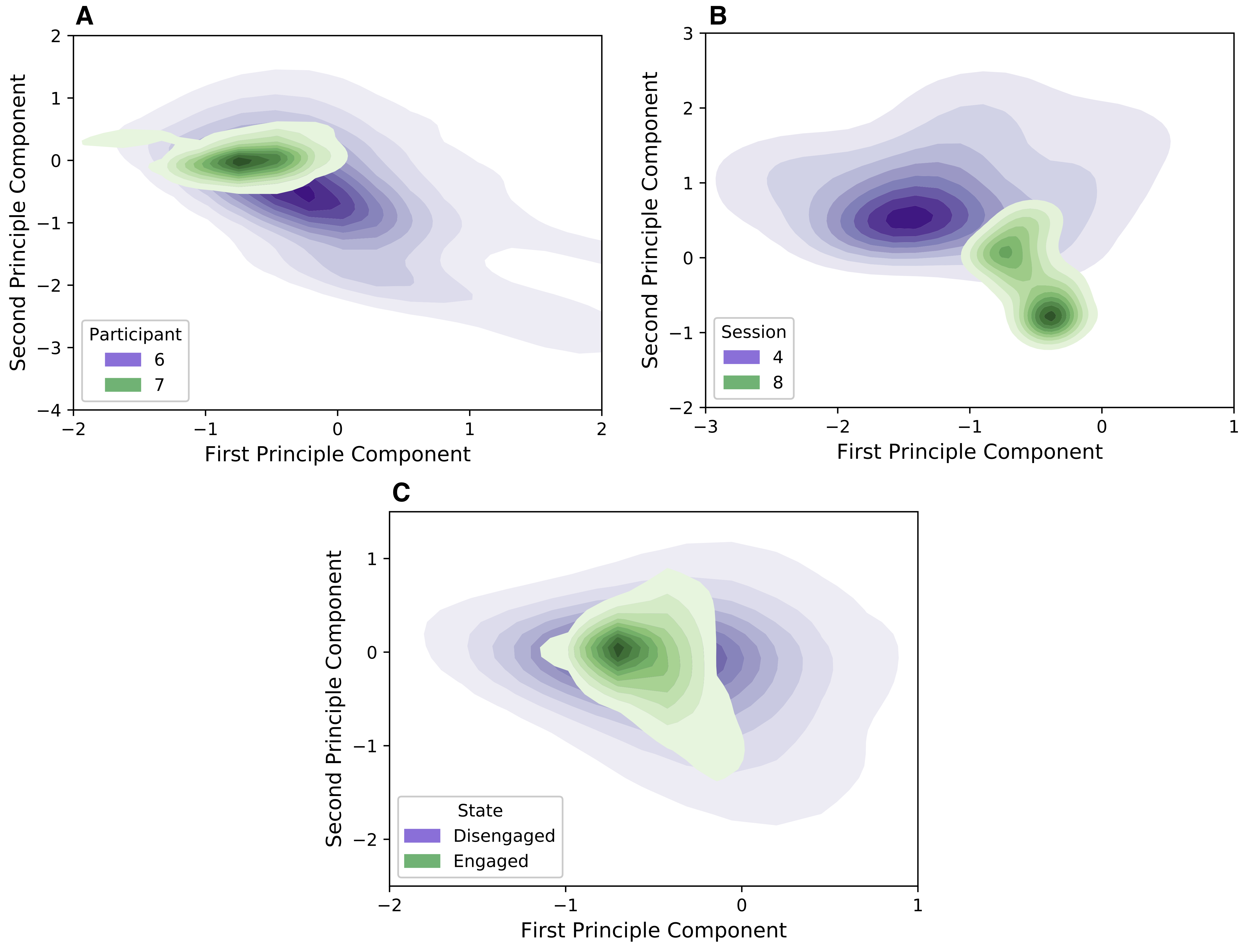}
    \caption{\bf{Comparing data across participants, sessions, and engagement states.} \normalfont{A visualization of limited overlap in data of two participants (}\bf{A}\normalfont{), two sessions from the same user (}\bf{B}\normalfont{), and across engagement states (}\bf{C}\normalfont{). Higher variance in data also observed when users are disengaged (}\bf{C}\normalfont{). Visualized data are those with high face detection confidence, compressed to two dimensions using principal component analysis (PCA); statistically significant ($p<0.01$) differences in means and variances determined using the complete dataset.}}
    \label{feature_map}
\end{figure}

This work's long-term, real-world setting resulted in significantly different means and variances of data across participants, sessions, and engagement states, as shown in Figure \ref{feature_map}. The figure compresses recorded data with high face detection confidence to two dimensions using principal component analysis (PCA), a commonly used unsupervised dimensionality reduction technique. Plotting compressed data reveals limited overlap between two participants (Figure \ref{feature_map}A) and two sessions from the same participant (Figure \ref{feature_map}B). Additionally, Figure \ref{feature_map}C shows a higher variance in data when participants are disengaged, which may explain the low recall values for disengagement reported in the previous section. Supplementary Figures \ref{sup_comparing_users} and \ref{sup_comparing_sessions} show similar visualizations for all participants and all sessions for the same participant in Figure \ref{feature_map}B.

Statistical analysis confirmed that both the means and variances of features differed significantly ($p<0.01$) across participants, sessions, and engagement states. We used a one-way analysis of variance (ANOVA) to test differences in means, and an F-test for differences in variance. Tests were performed on the principal components of all data.

\begin{figure}[t!] 
    \centering
    \includegraphics[width=0.65\linewidth]{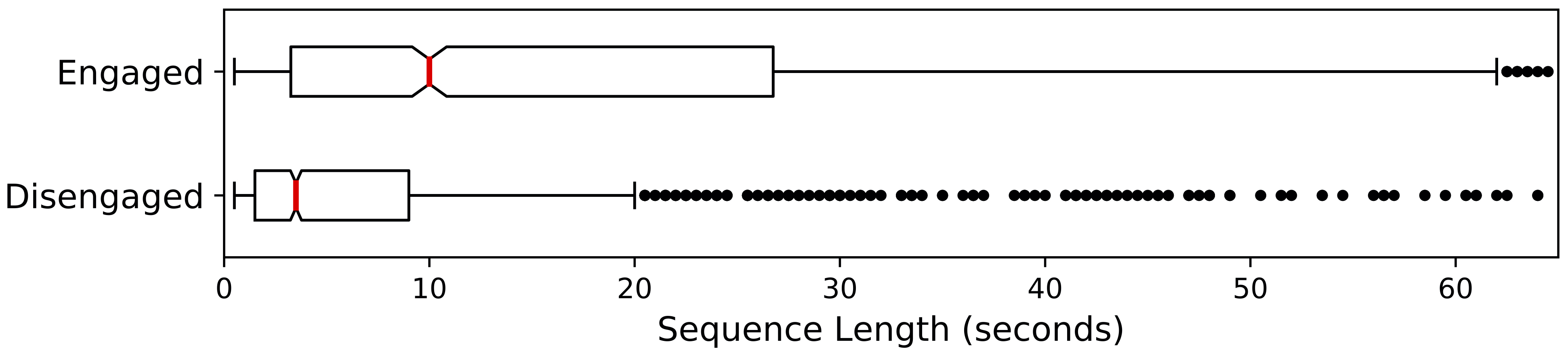}
    \caption{\bf{Sequences of engagement and disengagement.}
    \normalfont{Median duration of engagement sequences (11.0s) is significantly ($p<0.01$) higher than the median duration of disengagement sequences (4.0s). Despite prevalence of shorter sequences, disengagement sequences longer than the upper quartile (9.5s) accounted for 75\% of the total time users were disengaged.}
    } 
    \label{box_plot}
\end{figure}

\begin{figure}[t!] 
    \centering
    \includegraphics[width=0.7\linewidth]{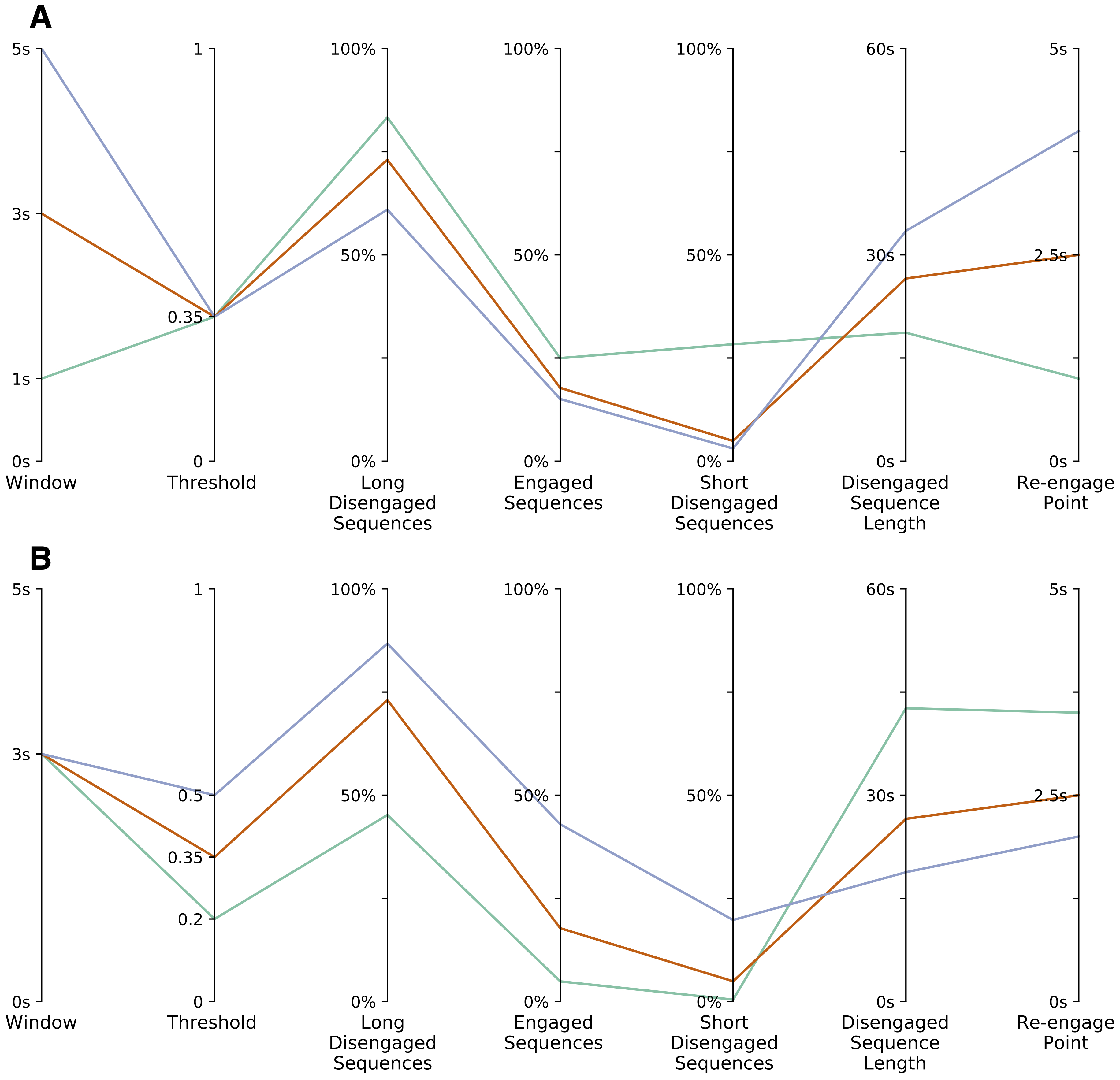}
    \caption{\bf{Re-engagement strategy results.}
    \normalfont{Post hoc analysis of strategy to re-engage users if predicted engagement probability is less than a threshold on average over a window. As shown by the orange line in }(\bf{A})\normalfont{ and }(\bf{B})\normalfont{, a threshold of 0.35 and window of 3s would have re-engaged users in 73\% of long disengagement sequences but also 15\% of engagement sequences. Varying window lengths }(\bf{A})\normalfont{ and thresholds }(\bf{B})\normalfont{ illustrates the trade-off between maximizing re-engagement in disengaged sequences and minimizing re-engagement in engaged sequences. Results based on generalized models trained on six participants.}
    } 
    \label{reeng_results}
\end{figure}

\paragraph*{Detecting Disengagement Sequences Using Temporal Patterns\newline}

This study demonstrates the importance and feasibility of detecting longer sequences of disengagement using the temporal structure of model predictions. {\it Engagement sequences} (ES) are periods in the interaction when the user is continuously engaged, while {\it disengagement sequences} (DS) are periods when the user is continuously disengaged. As Figure \ref{box_plot} shows, the duration of ES had an interquartile range of 5.0 to 27.0 seconds (s) while the duration of DS had an interquartile range of 2.5s to 9.5s. This work defines {\it long DS} as having a duration greater than the upper quartile (9.5s) and {\it short DS} as having a duration less than the lower quartile (2.5s). Long DS accounted for 75\% of the total time users were disengaged, whereas short DS accounted for only 5\% of the total time users were disengaged. This suggests that re-engagement strategies should focus on long DS despite the presence of many shorter sequences. 

The results and insights from the data suggest the following strategy for determining when to initiate re-engagement actions (RA): (i) average a model's predicted probability of engagement over a given window, and then (ii) initiate RA if the engagement probability is less than a given threshold. This approach should maximize long DS with RA, and minimize the percentage of ES with RA. Other considerations include the percentage of short DS with RA, the median duration of DS with RA, and the median elapsed time in DS before RA. The window length and threshold will affect these evaluation metrics so the choices for these parameters should depend on the intervention design and implemented RA.

Figure \ref{reeng_results} presents a post hoc analysis of the proposed re-engagement strategy. For example, suppose this study initiated RA if the predicted engagement probability was less than 0.35 on average for a 3s window. This approach would have led to RA in 73\% of long DS. The median duration of DS with RA would have been 25s, and RA would have occurred 2.5s into these sequences. However, RA would also have occurred in 5\% of short DS and 15\% of ES. 

Varying the window lengths and thresholds highlights the trade-off between maximizing RA in DS and minimizing RA in ES. The window length was negatively correlated with the percentage of long DS ($r_s=-0.74$) and ES ($r_s=-0.88$) with RA for a fixed threshold, as shown in Figure \ref{reeng_results}A. On the other hand, the threshold was positively correlated with the percentage of long DS ($r_s=+1.00$) and ES ($r_s=+1.00$) with RA for a fixed window length, as shown in Figure \ref{reeng_results}B. The reported results are based on generalized models trained on six users. Supplementary Tables \ref{sup_reeng_gen} and \ref{sup_reeng_ind} contain results for both generalized and individualized models with additional window length and threshold combinations. 

\paragraph*{Different Modalities and Model Types\newline}

\begin{figure}[t!] 
    \centering
    \includegraphics[width=\linewidth]{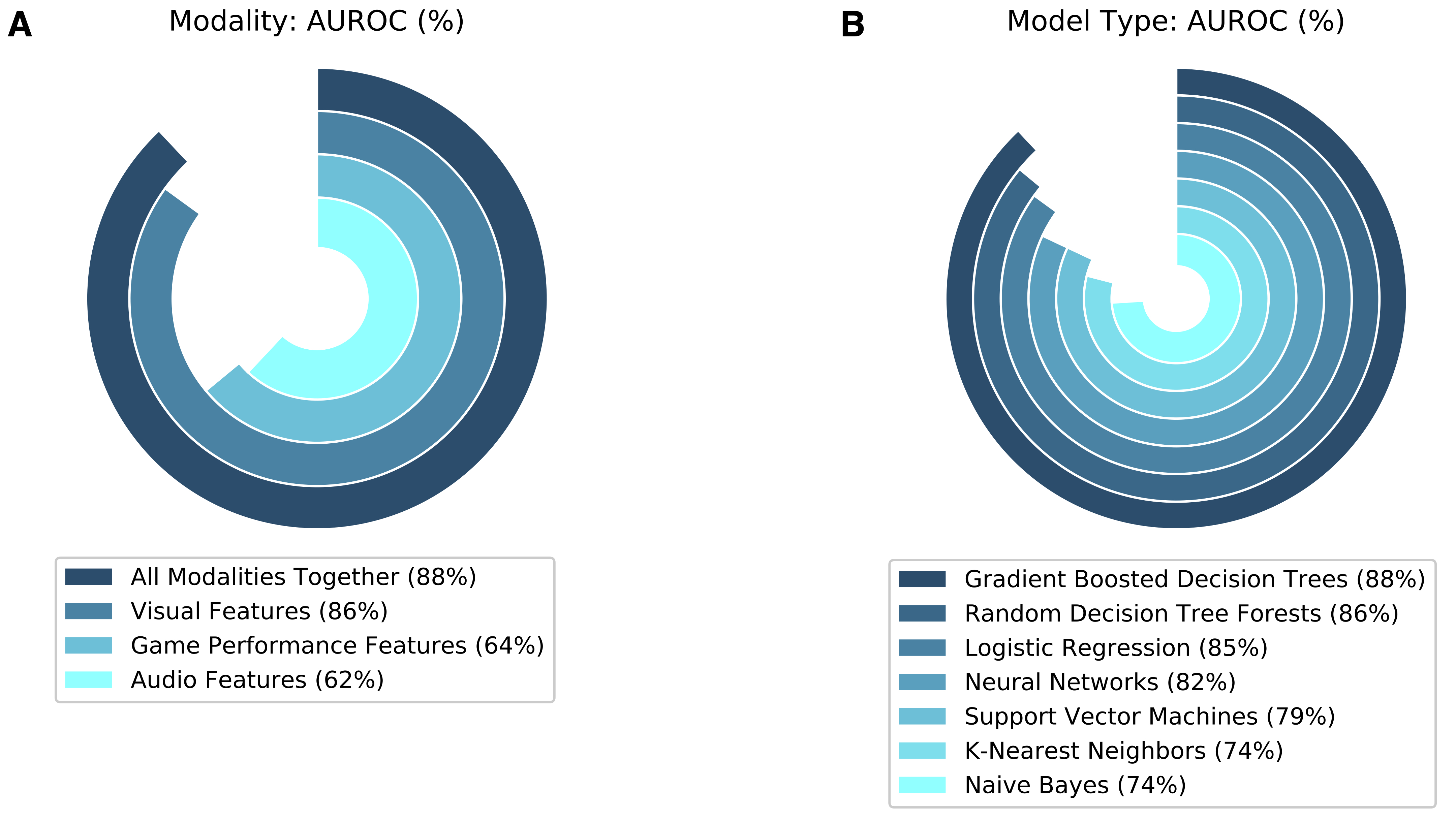}
    \caption{\bf{Results across different modalities and model types.} 
    \normalfont{All modalities together outperformed each modality separately, but visual features were the most significant (}\bf{A}\normalfont{). Tree-based models were the most successful among conventional supervised ML model types (}\bf{B}\normalfont{).}} 
    \label{comparing_results}
\end{figure}

Over the month-long, in-home SAR interventions, we collected a rich multimodal dataset from which we derived visual, audio, and game performance features to model engagement. To assess each modality's importance, we created separate models using each feature group. As Figure \ref{comparing_results}A shows, all modalities together outperformed each individual modality. However, models created using only visual features outperformed those created using audio or game performance features by about 20\% AUROC. These results support related work in this journal \cite{rudovic_science} that also found visual features as the most significant but multiple modalities as complimentary. 

Moreover, analyzing individual features revealed that the results of this work could largely be replicated using only seven key features. Feature analysis was performed using Pearson's correlation coefficient ($r$), and key features were determined using a threshold of $|r| > 0.20$. The key features are: the elapsed time in a session, the number of people in the environment, the direction of the user's eye gaze, the distance from the camera to the user, the elapsed time since the robot last talked, the count of incorrect responses to game questions, and the confidence value with which the user's face is being detected in the camera frame. Models using only these seven key features achieved AUROC values within 5\% of the results described above. 

Additionally, this work explored several supervised ML model types, but found tree-based models to be the most successful. The following conventional model types were considered: Naive Bayes, K-Nearest Neighbors, Support Vector Machines, Neural Networks, Logistic Regression, Random Decision Tree Forests, and Gradient Boosted Decision Trees. Of these, Gradient Boosted Decision Trees had the highest AUROC, as shown in Figure \ref{comparing_results}B, and are the basis for model evaluation metrics reported in previous sections. We also explored sequential models, such as Hidden Markov Models, Conditional Random Fields, and Recurrent Neural Networks, but found these to be less effective than conventional static models.

\newpage

A few alternative modeling approaches were considered as well. Interestingly, an ensemble of generalized and individualized models did not lead to better results than those approaches applied separately. Other explored approaches included: (i) rule-based models with the key features, (ii) deep neural networks with re-weighting techniques \cite{mengye_uber_approach}, and (iii) synthetic oversampling of the disengaged class \cite{bowyer_smote}. None of these approaches outperformed Gradient Boosted Decision Trees. 

\newpage
\section*{Discussion}
Robot perception of user engagement is a crucial HRI capability previously unexplored in the context of long-term, in-home SAR interventions for children with ASD. This study is the first to model engagement in this complex setting, and it also differs from previous work by developing supervised ML models intended for real-world deployments. The discussion below focuses on how this work highlights the feasibility and challenges of online recognition and response to disengagement in long-term, in-home SAR contexts. These contributions aim to inform the design of more engaging and personalized HRI, and improve SAR's effectiveness in augmenting learning by children with ASD. 

\paragraph*{Feasibility of Online Closed-Loop Implementation\newline}

This study presents supervised ML models that are feasible for use in online robot perception of and closed-loop response to disengagement. We developed two types of models for each user: (i) generalized models trained on data from different users; and (ii) individualized models trained on data from early subsets of users' interventions. The visual, audio, and performance features along with the labeled training data used in these models can be obtained in online deployments, as discussed further in Materials and Methods. Generalized and individualized models achieved approximately 90\% AUROC in this work's post hoc analysis. Individualized models performed better with additional data, likely because participants had higher engagement in early subsets used for training. Overall, the similar performance of these approaches indicates the possibility of having one model for multiple users. Generalized and individualized models also attained comparable AUROC to models trained on random samples of all participants' data, which is an ideal but impractical method. 

A shortcoming of generalized and individualized models is revealed through a 50\% false negative rate for detecting disengagement. This effect is likely due to the higher variance in features when participants were disengaged compared to when they were engaged. Despite the low recall values, a SAR system that accurately recognizes some instances of disengagement can still considerably enhance the interaction by attempting to re-engage the user at appropriate times. Analyzing \textit{disengagement sequences} (DS), or periods in the interaction when the user is continuously disengaged, shows that 75\% of the total time participants were disengaged occurred during DS that were 10 seconds long or longer. Some examples of participant behavior during these long DS included playing with toys, interacting with siblings, and even abruptly leaving the intervention setting. Shorter DS typically involved brief shifts in participant focus to other aspects of the environment. Moreover, most long DS required caregivers to re-engage participants, whereas participants re-engaged on their own in short DS. This suggests that re-engagement strategies should focus on counteracting longer DS. 

This work's post hoc analysis shows generalized and individualized models could be used to reliably initiate re-engagement actions (RA) during long DS. The presented re-engagement strategy would have initiated RA if the average predicted engagement probability over a window of time fell under a set threshold. Using a 0.35 threshold and a 3-second window would have resulted in RA for about 75\% of long DS, with the first RA occurring 2.5 seconds into these sequences on average; however, RA would also have been erroneously initiated in 15\% of engagement sequences (ES). An exploration of various window lengths and thresholds reveals the trade-off between maximizing RA in DS and minimizing RA in ES. This balance is important for maintaining RA effectiveness, and choices for these parameters should depend on the implemented RA and overall intervention design. 

The presented models are also readily interpretable, an important characteristic for facilitating implementation. Interpretability of ML is especially important in the ASD context, where therapists and caregivers need an understanding of the system's behavior in order to trust and adopt it \cite{rudovic_science}. The described models achieved interpretability in two ways: (i) through a simplified feature set and (ii) through the selected model types. First, this work replicated the described model accuracies to within 5\% using seven key features, as described in Results. This shows the problem of modeling engagement to be more tractable and provides insights for the design of future HRI studies in similar contexts. Second, we found tree-based ML models to be the most effective. Such methods are comparatively easier to train and interpret compared to more complex ML models \cite{li2019_interpretable}. Nevertheless, it is unknown whether the effectiveness of tree-based models in this work would generalize to other contexts. The interpretability of this work further demonstrates the feasibility of applying supervised ML to model user engagement online in closed-loop HRI. 

\paragraph*{Challenges of the Real-World In-Home Context\newline}

A long-term, real-world HRI setting raises many modeling challenges due to the significant noise and variance in data. The unconstrained home environment in particular presented several unforeseen problems. First, the camera was attached to the top of the game tablet, but it's position was frequently disturbed by both caregivers and child participants. For example, some caregivers temporarily turned the camera towards the ground or toward the robot when child participants were taking a break, instead of using the system's power switch as instructed. As a result, the camera position varied throughout the study adding noise to the extracted visual features. The audio data in this study also contained a high level of background noise, including sounds from television, pets, kitchen appliances, and lawn mowers. All participating families chose to place the SAR system in their living rooms, so external parties such as siblings, friends, and neighbors regularly interrupted sessions. The camera sometimes failed to capture all individuals in the environment, as the system was not designed for multi-party interactions. Finally, the variance in data was higher in this study given participants were children with ASD, who display atypical and highly diverse behaviors \cite{rudovic_science}. 

Substantial variance in data leaves supervised ML models vulnerable to overfitting. To mitigate this risk, standard ML practices such as bagging, boosting, and early stopping were used, as discussed below in Materials and Methods. In spite of the challenges of a real-world setting, generalized and individualized models achieved AUROC values around 90\% and could reliably initiate re-engagement actions in long sequences of disengagement. However, these models only had 50\% recall for disengagement. Further improving the false negative rate is a key area of future work. 

\paragraph*{Limitations and Future Work\newline}
As discussed in the previous section, a key modeling challenge in real-world HRI is the substantially increased variance in data, especially when users are disengaged. The solution to this problem in traditional ML is to obtain a large sample of labeled training data. This is not always feasible in HRI and is especially complex for atypical user populations. Moreover, this challenge is especially acute in the ASD context, where high variance in behaviors manifest not only between individuals but also within each individual. 

Active learning (AL) is a promising approach to this challenge, as it seeks to automatically select the most informative instances that need labeling \cite{clabaugh_AR_AL}. Preliminary work has shown AL to successfully train models of user engagement with a small amount of data \cite{rudovic_al}. However, AL is yet to be validated in long-term, real-world settings, as discussed previously in this journal \cite{clabaugh_SR_AL}. A future approach could first use supervised ML to train base models on available labeled data from other users or a user's beginning sessions, as done in this work. Then, AL could be applied to decide when to request a label for unseen data. A therapist or caregiver could provide the labels off-line after sessions, allowing the model to iteratively improve in a long-term setting.  

Ultimately, the most important direction for future work is to deploy ML frameworks online in real-world HRI and SAR. Such deployments are critical for understanding how well models recognize disengagement under realistic constraints of noise, uncertainty, and variance in data. When implemented online, these models could inform the activation of robot re-engagement actions; specifically, these could entail verbal and non-verbal robot responses such as socially supportive gestures and phrases \cite{brown_reengagement}. Overall, online recognition of and response to disengagement will enable the design of more engaging, personalized, and effective HRI, especially in SAR for the ASD community. 

\section*{Materials and Methods}

\paragraph*{Multimodal Dataset\newline}

The presented engagement models are based on data from month-long, in-home SAR interventions with children with ASD. During child-robot interactions, participants played a set of space-themed math games on a touchscreen tablet while a 6 degrees of freedom Stewart-platform robot named Kiwi provided verbal and expressive feedback, as shown in Figure \ref{interventions}. The study was approved by the Institutional Review Board of the University of Southern California (UP-16-0075(v), and we obtained informed consent from the children's caregivers. The 7 child participants in this work had a clinical diagnosis of ASD in mild to moderate ranges as described in the Diagnostic and Statistical Manual of Mental Disorders \cite{american2013diagnostic}. Supplementary Table \ref{sup_part_demo} reports the ages and genders of the participants: ages were between 3 years, 11 months and 7 years, 2 months; 3 were female and 4 were male. An earlier article provides further details about the SAR system and intervention design, with a focus on how the robot's feedback and instruction challenge levels were personalized to each user's unique learning patterns using online reinforcement learning \cite{mahajan_frontiers}.

Over the course of the month-long, in-home study, we collected a large multimodal dataset from which we derived visual, audio, and game performance features to model engagement. Due to numerous technological challenges commonly encountered in noisy real-world studies, this work only considers approximately 21 hours of interaction from 7 participants who had sufficient multimodal data. Participant 4 had the maximum interaction time analyzed (3 hours, 48 minutes), and Participant 6 had the minimum interaction time analyzed (1 hour, 52 minutes). Data collected in individual sessions were truncated to only use the content between the first and last game, because session data often included unstructured interactions before and after the games. Each participant was given a tutorial session as an introduction to the SAR system; data from that session were not included in the analysis. 

A USB camera mounted at the top of the game tablet recorded a front view of the participants. Visual and audio features were extracted from this camera data using OpenFace \cite{baltrusaitis_openface}, OpenPose \cite{cao_openpose}, and Praat \cite{boersma_praat}, open-source data processing tools feasible for online use in HRI. Visual features derived from OpenFace included: (i) the face detection confidence value, (ii) eye gaze direction, (iii) head position; and (iv) facial expression features. OpenPose was only used to estimate the number of people in the environment, since the camera's field of view centered on the user's face. Audio features derived from Praat included pitch, frequency, intensity, and harmonicity. Game performance features were also derived from system recordings and included the challenge level of the game being played, the count of incorrect responses to game questions, and the elapsed time in a session, game, and since the robot last talked. Supplementary Note S1 lists all visual, audio, and game performance features used for modeling engagement. 

In this work, a participant was {\it annotated to be engaged or disengaged} using standard definitions of engagement as a combination of behavioral, affective, and cognitive constructs \cite{scassellati_matric_review}. Specifically, a participant was considered to be engaged when paying full attention to the interaction, immediately responding to the robot's prompts, or seeking further guidance from others in the room. The lead author of this work annotated whether a participant was engaged or disengaged for the seven participants. To verify the absence of bias, two additional annotators independently annotated 20\% of the data for each participant; inter-rater reliability was measured using Fleiss' Kappa, and a reliability of $k=0.84$ was achieved between the primary and verifying annotators. Supplementary Table \ref{sup_anno_crit} contains the specific criteria followed by all annotators. 

\paragraph*{Modeling Approaches\newline}

This work applied and evaluated conventional supervised ML methods to model user engagement in month-long, in-home SAR interventions for children with ASD. First, we applied a few preprocessing techniques to the multimodal dataset described in the previous section to address missing data and possible errors in the fusion of modalities. Although we obtained data for each camera frame at a standard rate of 30 frames per second, this work considered the median value of features and annotations in overlapping one second intervals (i.e., 0 to 1 second, 0.5 to 1.5 seconds, 1 to 2 seconds, etc.). The following features were added for each interval: (i) the variance of continuous-valued features in the interval; and (ii) an indicator for whether discrete-valued features changed in the interval. This also addressed the problem of low OpenFace confidence for detecting the user's face; low confidence occurred in 38\% of camera frames overall, but in only 3 continuous frames on average. Furthermore, all features were standardized to have zero mean and unit variance since raw values were measured on different scales. The means and variances of each feature needed for standardization were obtained with respect to the training set in order to maintain the feasibility of online implementation. 

To model user engagement, this work used two supervised ML approaches that are practical for online implementation in closed-loop HRI: (i) \textit{generalized models} trained on data from different users; and (ii) \textit{individualized models} trained on data from early subsets of the users' interventions. Generalized models were implemented by training on data from a given subset of $M$ participants. The models were then tested on the remaining $N$ users not in the training subset. We considered all possible combinations of participants; because there were 7 participants, values of $M$ and $N$ ranged from 1 to 6. Individualized models were developed by sorting a user's data chronologically and using an early subset for training and later subset for testing the model. In particular, we applied this evaluation to training sets from the first 10\% to the first 90\% of a user's data, in increments of 10\%. We used this approach to standardize the training sets across differences in participant interaction times; future implementations could use beginning sessions as training data instead. The generalized and individualized approaches are both feasible for online use in HRI deployments; the labeled training data required for supervised ML models can be obtained in both cases. In order to evaluate these approaches, we also implemented models trained and tested on distinct random samples of all users' data despite this approach being impractical for online use. The proportions of training data evaluated in the random sampling approach also ranged from 10\% to 90\%, in increments of 10\%.

Using the generalized, individualized, and random sampling approaches, this work implemented several supervised ML model types. All considered model types are reported in the Results section; Gradient Boosted Decision Trees were the most successful. Specifically, we implemented Gradient Boosted Decision Trees with early stopping and bagging \cite{xgboost}. Boosting algorithms train weak learners sequentially, with each learner trying to correct its predecessor. Early stopping partitions a small percentage of the training data for validation, and ends training when performance on the validation set stops improving. Bagging fits several base classifiers on random subsets of the training data, and then aggregates the individual predictions to form a final prediction. These techniques were adopted to mitigate the increased risk of overfitting in high variance datasets, as was the case in this long-term, in-home study. 

We implemented the ML models in Python using the following libraries: Scikit-learn version 0.21.3 \cite{scikit-learn}, XGBoost version 0.90 \cite{xgboost}, Hmmlearn version 0.2.1 \cite{hmmlearn}, CRFsuite version 0.12 \cite{crf_suite}, TensorFlow version 1.15.0 \cite{tensorflow}, and Keras version 2.2.4 \cite{keras}. All models were implemented with default hyperparameters, as specified in Supplementary Table \ref{sup_model_hp}. We used default hyperparameters because the variance in data caused commonly used strategies such as cross validation and grid search to overfit to the training data. All reported model results are from Scikit-learn implementations, except for Gradient Boosted Decision Trees, which we implemented using XGBoost for improved computational performance. Neural Networks were also implemented in TensorFlow and Keras, and had similar performance to the reported results from Scikit-learn. Sequential models were explored using Hmmlearn, CRFsuite, and Keras. 

\clearpage

\section*{Acknowledgments}

The authors thank K. Mahajan and L. Mathur for their help with data analysis, K. Peng and J. Keller for their assistance with annotations, and T. Groechel, L. Klein, and R. Pakkar for their advice and support. In addition, the authors are very grateful to G. Ragusa for a key role in the original study design, recruitment, assessments, and more. The entire research team thanks the children and families who participated in the study that generated the dataset.

\paragraph{Funding:} This research was supported by the National Science Foundation Expedition in Computing Grant NSF IIS-1139148. 

\paragraph{Author contributions:} S.J. led this work's conceptualization and investigation. S.J., B.T., and Z.S. processed the data, implemented the methods, and analyzed the results. C.C. designed the study and managed the deployments that generated the datasets used in this work. M.J.M. was the leading faculty advisor for the overarching study on SAR for ASD, and all conducted research, data collection, and analysis. S.J., B.T., Z.S., and M.J.M. all contributed to the writing of this article. 

\paragraph{Competing interests:} M.J.M. is a co-founder of Embodied, Inc. but has not been involved with the company since December 2016. C.C. is now a full-time employee of Embodied, Inc. but was not involved with the company while the reported work was done.

\paragraph{Data and materials availability:} All data needed to evaluate the conclusions that can be released under USC IRB policies are included in the article or the Supplementary Materials. Please contact S.J. and M.J.M. for questions about other materials. 

\clearpage

\section*{Supplementary Materials}

\begin{figure}[h!] 
    \centering
    \includegraphics[width=0.8\linewidth]{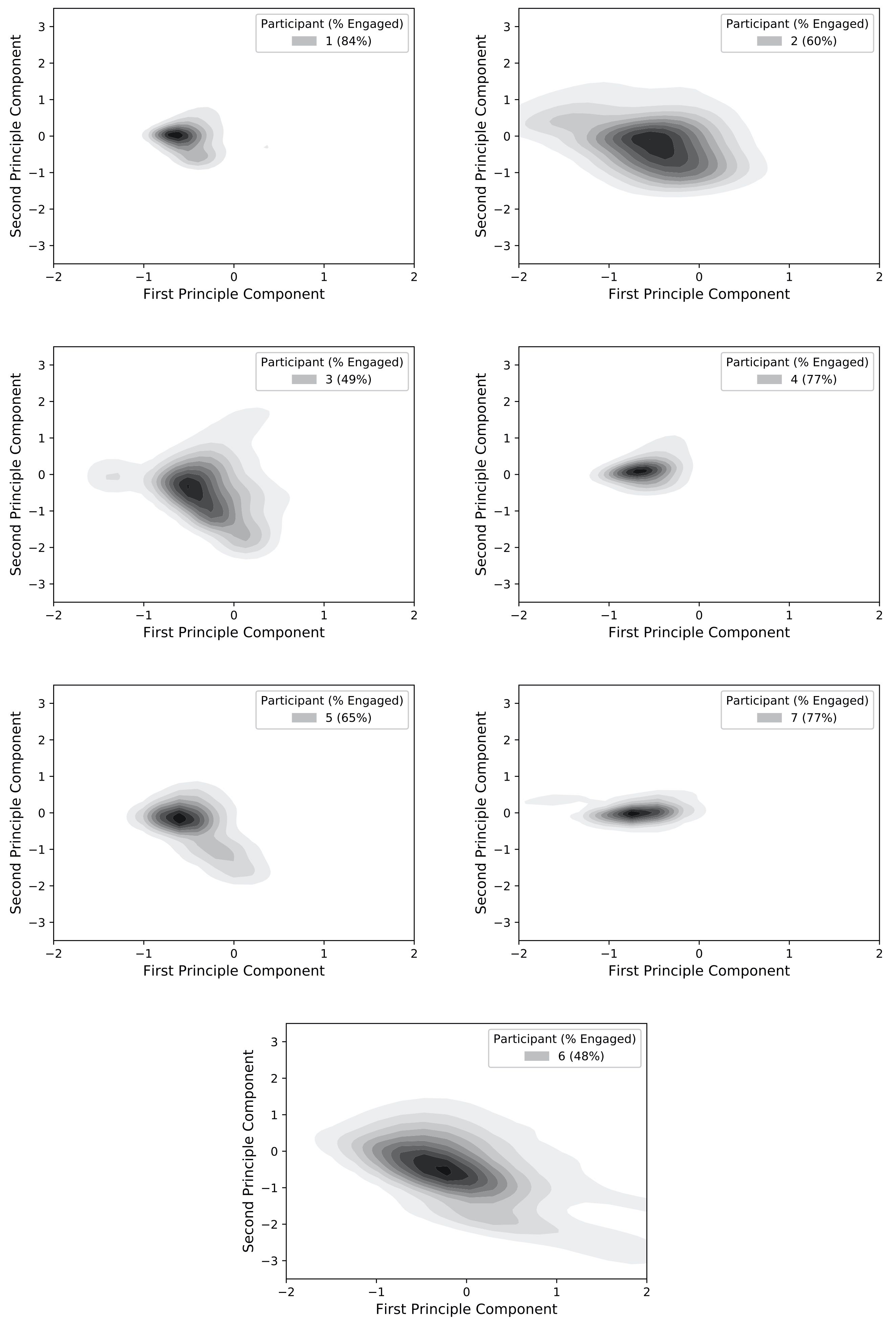}
    \caption{\bf{Comparing data across users.} 
    \normalfont{Statistically significant ($p<0.01$) difference in means and variances of all recorded data across users. Visualized data are those with high face detection confidence, compressed to two dimensions using principal component analysis (PCA).}} 
    \label{sup_comparing_users}
\end{figure}

\begin{figure}[h!] 
    \centering
    \includegraphics[width=0.8\linewidth]{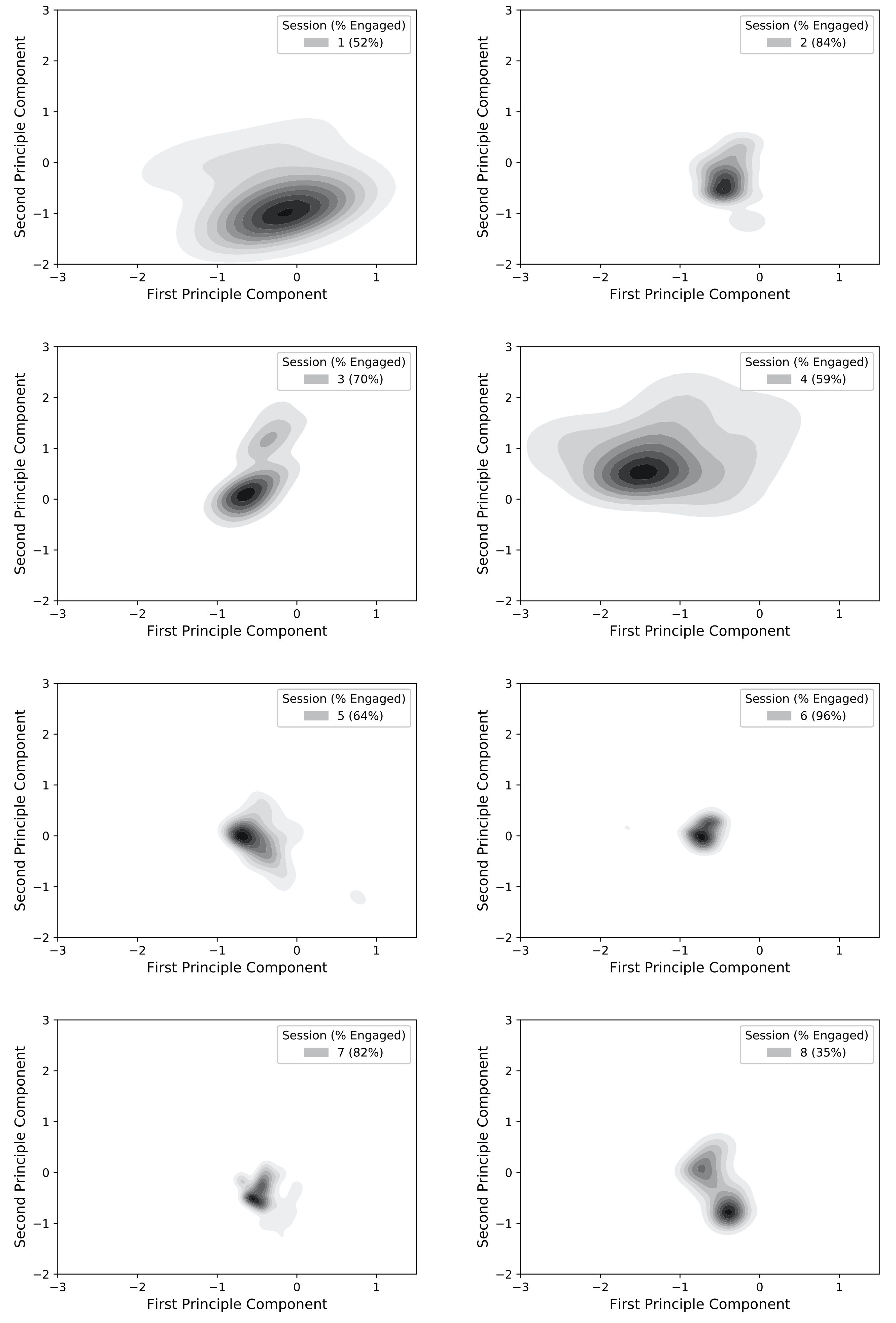}
    \caption{\bf{Comparing data across sessions.} 
    \normalfont{Statistically significant ($p<0.01$) difference in means and variances of all recorded data across sessions for each user. Sessions shown here from Participant two. Visualized data are those with high face detection confidence, compressed to two dimensions using principal component analysis (PCA).}} 
    \label{sup_comparing_sessions}
\end{figure}

File S1 (.csv format). Dataset for engagement models. \href{https://drive.google.com/file/d/1LbP7UVoI8B4LDp_eD085m21RqGzxkuA0/view?usp=sharing}{\textit{Google Drive Link.}}\newline 
File S2 (.csv format). Descriptions of columns in File S1. \href{https://drive.google.com/open?id=1aOo7UsE4aQLnWyl8DcM-Ml6kMrsIcnF5}{\textit{Google Drive Link.}} \newline  
Note S1. List of multimodal features.
\newline
Figure \ref{sup_comparing_users}. Comparing data across users.
\newline
Figure \ref{sup_comparing_sessions}. Comparing data across sessions.
\newline
Table \ref{sup_part_demo}. Participant demographic information.
\newline
Table \ref{sup_anno_crit}. Engagement annotation criteria.
\newline
Table \ref{sup_model_hp}. Model hyperparameters.
\newline
Table \ref{sup_gen_results}. Generalized model results.
\newline
Table \ref{sup_ind_results}. Individualized model results.
\newline
Table \ref{sup_rand_results}. Random sampling model results.
\newline
Table \ref{sup_reeng_gen}. Re-engagement strategy evaluation using generalized models.
\newline
Table \ref{sup_reeng_ind}. Re-engagement strategy evaluation using individualized models.

\clearpage

\bibliographystyle{science_bib}
\bibliography{bibliography} 

\clearpage


\beginsupplement

\section*{Supplementary Materials}

\paragraph{Note S1. List of multimodal features.\newline}

This work is based on a large multimodal dataset from month-long, in-home SAR interventions for children with ASD. Video and audio features were extracted from camera data using OpenFace \cite{baltrusaitis_openface}, OpenPose \cite{cao_openpose}, and Praat \cite{boersma_praat}. Game performance features were derived from the tablet interactions. A list of all features used for modeling engagement is included here. 

\begin{itemize}
  \item Visual Features :
  \begin{itemize}
      \item Face Detection: (i) confidence value for face detection, (ii) binary success value for face detection, and (iii) elapsed time since last successful detection;
      \item Eye Gaze: (i) eye gaze direction vector for left eye, right eye, and both eyes (average), and (ii) Euclidean distance from camera to endpoint of participant's gaze (using intersection of gaze with vertical camera plane);
      \item Head Position: (i) orientation of head in terms of pitch, roll, and yaw movements, and (ii) Euclidean distance from camera to location of the participant's head;
      \item Facial Expression: The following OpenFace Action Units \cite{baltrusaitis_of_action_units} were used: (i) inner brow raiser, (ii) outer brow raiser, (iii) brow lowerer, (iv) upper lid raiser, (v) cheek raiser, (vi) lid tightener, (vii) nose wrinkler, (viii) upper lip raiser, (ix) lip corner puller, (x) dimpler, (xi) lip corner depressor, (xii) chin raiser, (xiii) lip stretcher, (xiv) lip tightener, (xv) lips part, (xvi) jaw drop, (xvii) lip suck, and (xviii) blink;
      \item Environment: (i) estimated number of people in the environment.
  \end{itemize}
  \item Audio Features: (i) harmonicity: measure of voice quality, (ii) intensity: power carried by sound waves per unit area, (iii) frequency: Mel-frequency cepstral coefficients, and (iv) pitch frequency and periodicity.
  \item Game Performance Features: (i) challenge level of game being played, (ii) count of incorrect responses to game questions in current game and overall session, (iii) number of games played in the session, (iv) game type, and (v) elapsed time in a session, game, and since the robot last talked.
\end{itemize}

\clearpage

\clearpage

\begin{table}[h!]
\centering
\begin{tabular}{|p{5cm}|p{5cm}|} 
\hline
Gender & Age \\ 
\hline\hline
Male & 4 years, 6 months \\
\hline
Female & 4 years, 11 months \\
\hline
Male & 4 years, 5 months \\
\hline
Male & 7 years, 2 months \\
\hline
Female & 3 years, 11 months \\
\hline
Male & 4 years, 6 months \\
\hline
Female & 7 years, 2 months \\ 
\hline
\end{tabular}
\caption{\bf{Participant demographic information.} \normalfont{Gender and age of the 7 child participants, in random order. All participants had a clinical diagnosis of ASD in mild to moderate ranges as described in the Diagnostic and Statistical Manual of Mental Disorders \cite{american2013diagnostic}. Additional details about the participants and study that generated the dataset are reported in an earlier article \cite{mahajan_frontiers}.}} \label{sup_part_demo}
\end{table}

\begin{table}[h!]
\centering
\begin{tabular}{|p{7cm}|p{7cm}|} 
\hline
Engagement & Disengagement \\ 
\hline\hline
paying full attention to the interaction
& not paying full attention to the interaction \\ 
\hline
present in the camera frame and in front of the system & not present in the camera frame or away from the system \\ [2.0ex]
\hline
attention directed toward the tablet or robot; listening or responding to the robot's prompts, interacting with the tablet
& when prompted, not listening or responding to the robot, not interacting with the tablet \\
\hline
looks or turns away from the system for further guidance or feedback from others in the room
& looks or turns away from the system aimlessly or because of a distraction, not for help
\\
\hline
very interested with minimal or no incentive from others in the room & only interested if prompted by others in the room or unresponsive to others in the room
\\ 
\hline
talking or arguing about the games & talking or arguing not about the games \\
\hline
\end{tabular}
\caption{\bf{Engagement annotation criteria.} \normalfont{Participants were annotated as engaged or disengaged in each camera frame using the criteria above.}} \label{sup_anno_crit}
\end{table}

\begin{table}[h!]
\centering
\begin{tabular}{|p{4cm}|p{4.1cm}|p{6.5cm}|}
\hline
Model & Hyperparameter & Description \\
\hline
\hline
\multirow{2}{4cm}{Naive Bayes} & type = Gaussian & assume Gaussian distributed features  \\
\cline{2-3}
& var\_smoothing = 1e-09 & variance proportion added for stability \\
\hline
\multirow{2}{4cm}{K-Nearest Neighbors} & n\_neighbors = 5 & number of neighbors considered \\
\cline{2-3}
& weights = uniform & neighbors are weighted equally \\
\hline
\multirow{2}{4cm}{Support Vector \newline Machines} & penalty = l2 & L2 norm used for penalization \\
\cline{2-3}
& loss = squared\_hinge & square of hinge loss function \\
\hline
\multirow{2}{4cm}{Logistic Regression} & penalty = l2 & L2 norm used for penalization \\
\cline{2-3}
& solver = liblinear & algorithm for optimization problem \\
\hline
\multirow{3}{4cm}{Random Decision \newline Tree Forests} & n\_estimators = 100 & 100 decision trees in random forest \\
\cline{2-3}
& max\_depth = None & nodes expanded until all leaves pure \\
\cline{2-3}
& max\_features = None & all features considered for node split \\
\hline
\multirow{3}{4cm}{Gradient Boosted Decision Trees} & n\_estimators = 100 & 100 decision trees in aggregate \\
\cline{2-3}
& max\_depth = 6 & maximum depth of the tree \\
\cline{2-3}
& booster = gbtree & algorithm for gradient boosting \\
\hline
\multirow{3}{4cm}{Neural Networks} & hidden\_layer\_sizes=(100) & one hidden layer with 100 nodes  \\
\cline{2-3}
& activation = relu & ReLU activation function \\
\cline{2-3}
& solver = adam & algorithm for optimization problem \\
\hline
\multirow{3}{4cm}{Hidden Markov Models} & type = Gaussian & assume Gaussian emissions  \\
\cline{2-3}
& n\_components = 2 & 2 hidden states: engaged, disengaged \\
\cline{2-3}
& algorithm = viterbi & algorithm for optimization problem \\
\hline
\multirow{2}{4cm}{Conditional Random Fields} & solver = lbfgs & algorithm for optimization problem  \\
\cline{2-3}
& regularizer = elastic\_net & elastic net (L1 and L(ii) regularization \\
\hline
\multirow{3}{4cm}{Recurrent Neural Networks} & hidden\_layer\_sizes=(100) & one LSTM layer with 100 nodes  \\
\cline{2-3}
& activation = relu & ReLU activation function \\
\cline{2-3}
& solver = adam & algorithm for optimization problem \\
\hline
\end{tabular}
\caption{\bf{Model hyperparameters.} \normalfont{Several supervised ML model types were considered, as shown here and detailed in Results. All models were implemented using the default hyperparameters shown here because substantial variance in data caused hyperparameter tuning strategies to overfit to training sets. ML models were implemented in Python using the following libraries: Scikit-learn  \cite{scikit-learn}, XGBoost  \cite{xgboost}, Hmmlearn \cite{hmmlearn}, CRFsuite \cite{crf_suite}, TensorFlow  \cite{tensorflow}, and Keras \cite{keras}.}}
\label{sup_model_hp}
\end{table}

\begin{table}[b!]
\centering
\begin{tabular}{|p{1.75cm}||p{1.75cm}|p{1.75cm}|p{1.75cm}|p{1.75cm}|p{1.75cm}|p{1.75cm}|}
\hline
\multirow{2}{1.75cm}{Training Users} & \multirow{2}{*}{AUROC} & \multirow{2}{*}{Accuracy} & \multicolumn{2}{|l|}{Engagement} & \multicolumn{2}{|l|}{Disengagement} \\ 
\cline{4-7}
& & & Precision & Recall & Precision & Recall \\ 
\hline\hline
1 & 85\% & 78\% & 81\% & 86\% & 72\% & 61\% \\
\hline
2 & 87\% & 79\% & 82\% & 88\% & 74\% & 64\% \\
\hline
3 & 88\% & 80\% & 82\% & 89\% & 75\% & 64\% \\
\hline
4 & 88\% & 80\% & 82\% & 90\% & 76\% & 61\% \\
\hline
5 & 88\% & 80\% & 82\% & 90\% & 75\% & 62\% \\
\hline
6 & 88\% & 81\% & 81\% & 91\% & 72\% & 57\% \\
\hline
\end{tabular}
\caption{\bf{Generalized model results.} \normalfont{Model evaluation metrics for the generalized approach: training on a given subset of users and testing on different users. Evaluated for all possible combinations of the 7 participants, and results averaged across models with the same number of training users. Results reported for Gradient Boosted Decision Trees, the most successful model type.}}
\label{sup_gen_results}
\end{table}

\begin{table}[b!]
\begin{tabular}{|p{1.75cm}||p{1.75cm}|p{1.75cm}|p{1.75cm}|p{1.75cm}|p{1.75cm}|p{1.75cm}|}
\hline
\multirow{2}{1.75cm}{Training Proportion} & \multirow{2}{*}{AUROC} & \multirow{2}{*}{Accuracy} & \multicolumn{2}{|l|}{Engagement} & \multicolumn{2}{|l|}{Disengagement} \\ 
\cline{4-7}
& & & Precision & Recall & Precision & Recall \\ 
\hline\hline
10\% & 77\% & 72\% & 73\% & 90\% & 65\% & 33\% \\
\hline
20\% & 81\% & 76\% & 75\% & 93\% & 69\% & 39\% \\
\hline
30\% & 82\% & 77\% & 75\% & 94\% & 73\% & 39\% \\
\hline
40\% & 84\% & 76\% & 73\% & 93\% & 77\% & 41\% \\
\hline
50\% & 87\% & 78\% & 76\% & 95\% & 80\% & 49\% \\
\hline
60\% & 89\% & 79\% & 76\% & 95\% & 85\% & 49\% \\
\hline
70\% & 88\% & 79\% & 76\% & 95\% & 82\% & 45\% \\
\hline
80\% & 89\% & 82\% & 78\% & 95\% & 83\% & 52\% \\
\hline
90\% & 89\% & 82\% & 74\% & 92\% & 82\% & 58\% \\
\hline
\end{tabular}
\caption{\bf{Individualized model results.} \normalfont{Model evaluation metrics for the individualized approach: sorting a user's data chronologically, using an early subset for training and later subset for testing. Applied to proportions of training data ranging from the first 10\% to the first 90\% of a user's data, in increments of 10\%; results averaged across all users. Results reported for Gradient Boosted Decision Trees, the most successful model type.}}
\label{sup_ind_results}
\end{table}

\begin{table}[t!]
\centering
\begin{tabular}{|p{1.75cm}||p{1.75cm}|p{1.75cm}|p{1.75cm}|p{1.75cm}|p{1.75cm}|p{1.75cm}|}
\hline
\multirow{2}{1.75cm}{Training Proportion} & \multirow{2}{*}{AUROC} & \multirow{2}{*}{Accuracy} & \multicolumn{2}{|l|}{Engagement} & \multicolumn{2}{|l|}{Disengagement} \\ 
\cline{4-7}
& & & Precision & Recall & Precision & Recall \\ 
\hline\hline
10\% & 85\% & 87\% & 88\% & 93\% & 85\% & 77\% \\
\hline
20\% & 88\% & 90\% & 90\% & 94\% & 88\% & 81\% \\
\hline
30\% & 89\% & 91\% & 91\% & 95\% & 90\% & 83\% \\
\hline
40\% & 90\% & 92\% & 92\% & 95\% & 91\% & 84\% \\
\hline
50\% & 91\% & 92\% & 93\% & 96\% & 92\% & 86\% \\
\hline
60\% & 91\% & 93\% & 93\% & 96\% & 93\% & 87\% \\
\hline
70\% & 92\% & 93\% & 94\% & 96\% & 93\% & 87\% \\
\hline
80\% & 92\% & 94\% & 94\% & 97\% & 93\% & 88\% \\
\hline
90\% & 93\% & 94\% & 94\% & 97\% & 93\% & 89\% \\
\hline
\end{tabular}
\caption{\bf{Random sampling model results.} \normalfont{Model evaluation metrics for the random sampling approach: training and testing on distinct random samples of all users data. Applied to proportions of training data ranging from 10\% to 90\%, in increments of 10\%; results averaged over 10 iterations. Results reported for Gradient Boosted Decision Trees, the most successful model type.}}
\label{sup_rand_results}
\end{table}

\begin{table}[t!]
\centering
\begin{tabular}{|p{1.75cm}|p{1.75cm}|p{1.75cm}|p{1.75cm}|p{1.75cm}|p{1.75cm}|p{1.75cm}|} 
\hline
Window & Threshold & Long DS & ES & Short DS & DS Length & Re-engage Point \\ 
\hline\hline
3.0s & 0.10 & 17.8\% & 0.5\% & 0.2\% & 79.3s & 5.0s \\ 
\hline
3.0s & 0.15 & 33.6\% & 3.0\% & 0.5\% & 51.4s & 4.0s \\ 
\hline
3.0s & 0.20 & 45.2\% & 4.9\% & 0.5\% & 42.6s & 3.5s \\ 
\hline
3.0s & 0.25 & 55.5\% & 7.5\% & 1.2\% & 35.9s & 3.0s \\ 
\hline
3.0s & 0.30 & 64.4\% & 12.0\% & 2.1\% & 30.5s & 3.0s \\ 
\hline
3.0s & 0.35 & 73.1\% & 17.7\% & 4.9\% & 26.6s & 2.5s \\ 
\hline
3.0s & 0.40 & 77.6\% & 24.6\% & 8.6\% & 23.7s & 2.5s \\ 
\hline
3.0s & 0.45 & 82.4\% & 33.8\% & 12.4\% & 21.2s & 2.0s \\ 
\hline
3.0s & 0.50 & 86.8\% & 43.0\% & 19.8\% & 18.8s & 2.0s \\ 
\hline\hline
1.0s & 0.35 & 83.3\% & 25.0\% & 28.3\% & 18.7s & 1.0s \\ 
\hline
2.0s & 0.35 & 78.3\% & 19.7\% & 13.8\% & 22.3s & 2.0s \\ 
\hline
3.0s & 0.35 & 73.1\% & 17.7\% & 4.9\% & 26.6s & 2.5s \\ 
\hline
4.0s & 0.35 & 66.9\% & 15.9\% & 3.2\% & 30.0s & 3.5s \\ 
\hline
5.0s & 0.35 & 61.0\% & 15.1\% & 3.1\% & 33.5s & 4.0s \\ 
\hline
6.0s & 0.35 & 56.6\% & 13.6\% & 2.8\% & 36.2s & 5.0s \\ 
\hline
7.0s & 0.35 & 53.9\% & 12.6\% & 2.5\% & 39.3s & 5.5s \\ 
\hline
8.0s & 0.35 & 51.1\% & 11.9\% & 2.3\% & 42.4s & 6.5s \\ 
\hline
9.0s & 0.35 & 49.5\% & 11.1\% & 2.3\% & 45.2s & 7.0s \\ 
\hline
10.0s & 0.35 & 48.2\% & 10.8\% & 2.1\% & 47.3s & 7.5s \\ 
\hline
\end{tabular}
\caption{\bf{Re-engagement strategy evaluation using generalized models.} \normalfont{Post hoc strategy to re-engage users if predicted engagement probability is less than a threshold on average over a window. Analysis based on generalized models trained on 6 users for varying thresholds and window lengths. Evaluated on the following metrics: percentage of long disengagement sequences (DS) with RA, percentage of engagement sequences (ES) with RA, percentage of short DS with RA, the median duration of DS with RA, and the median elapsed time in DS before RA.}}
\label{sup_reeng_gen}
\end{table}

\begin{table}[t!]
\centering
\begin{tabular}{|p{1.75cm}|p{1.75cm}|p{1.75cm}|p{1.75cm}|p{1.75cm}|p{1.75cm}|p{1.75cm}|} 
\hline
Window & Threshold & Long DS & ES & Short DS & DS Length & Re-engage Point \\ 
\hline\hline
3.0s & 0.10 & 30.7\% & 1.6\% & 0.9\% & 46.1s & 3.0s \\ 
\hline
3.0s & 0.15 & 37.3\% & 4.6\% & 3.7\% & 37.4s & 2.5s \\ 
\hline
3.0s & 0.20 & 43.9\% & 7.1\% & 5.8\% & 31.6s & 2.5s \\ 
\hline
3.0s & 0.25 & 51.3\% & 10.4\% & 7.3\% & 28.0s & 2.5s \\ 
\hline
3.0s & 0.30 & 55.7\% & 14.6\% & 8.8\% & 26.1s & 2.5s \\ 
\hline
3.0s & 0.35 & 59.6\% & 18.2\% & 10.4\% & 24.0s & 2.0s \\ 
\hline
3.0s & 0.40 & 64.5\% & 22.6\% & 11.9\% & 22.8s & 2.0s \\ 
\hline
3.0s & 0.45 & 66.7\% & 27.2\% & 15.5\% & 20.8s & 2.0s \\ 
\hline
3.0s & 0.50 & 72.8\% & 30.8\% & 17.1\% & 19.4s & 2.0s \\ 
\hline\hline
1.0s & 0.35 & 70.6\% & 23.8\% & 28.0\% & 18.6s & 1.0s \\ 
\hline
2.0s & 0.35 & 64.9\% & 19.8\% & 17.7\% & 21.3s & 1.5s \\ 
\hline
3.0s & 0.35 & 59.6\% & 18.2\% & 10.4\% & 24.0s & 2.0 \\ 
\hline
4.0s & 0.35 & 56.1\% & 16.6\% & 7.9\% & 27.0s & 3.0s \\ 
\hline
5.0s & 0.35 & 53.9\% & 14.9\% & 7.0\% & 29.0s & 3.0s \\ 
\hline
6.0s & 0.35 & 50.0\% & 14.0\% & 6.7\% & 30.7s & 3.5s \\ 
\hline
7.0s & 0.35 & 48.2\% & 13.1\% & 7.3\% & 31.7s & 3.5s \\ 
\hline
8.0s & 0.35 & 46.9\% & 12.7\% & 7.0\% & 33.6s & 3.0s \\ 
\hline
9.0s & 0.35 & 44.7\% & 11.3\% & 7.0\% & 35.0s & 2.5s \\ 
\hline
10.0s & 0.35 & 41.7\% & 11.0\% & 6.7\% & 35.9s & 2.0s \\ 
\hline
\end{tabular}
\caption{\bf{Re-engagement strategy evaluation using individualized models.} \normalfont{Post hoc strategy to re-engage users if predicted engagement probability is less than a threshold on average over a window. Analysis based on individualized models trained on the first 50\% of users' data for varying thresholds and window lengths. Evaluated on the following metrics: percentage of long disengagement sequences (DS) with RA, percentage of engagement sequences (ES) with RA, percentage of short DS with RA, the median duration of DS with RA, and the median elapsed time in DS before RA.}}
\label{sup_reeng_ind}
\end{table}

\end{document}